\documentclass{PoS}

\title{Tensor Networks and their use for Lattice Gauge Theories}

\ShortTitle{Tensor Networks for LGT}

\author{\speaker{Mari Carmen Ba{\~n}uls}\\ 
Max-Planck-Institut f\"ur Quantenoptik, Hans-Kopfermann-Stra{\ss}e 1, 85748 Garching, Germany\\
        E-mail: \email{banulsm@mpq.mpg.de}}

\author{Krzysztof Cichy\\
Faculty of Physics, Adam Mickiewicz University, Umultowska 85, 61-614 Pozna\'{n}, Poland\\
E-mail: \email{krzysztof.cichy@gmail.com}}

\author{J. Ignacio Cirac\\
Max-Planck-Institut f\"ur Quantenoptik, Hans-Kopfermann-Stra{\ss}e 1, 85748 Garching, Germany\\
E-mail: \email{ignacio.cirac@mpq.mpg.de}}

\author{Karl Jansen\\
NIC, DESY Zeuthen, Platanenallee 6, 15738 Zeuthen, Germany\\
E-mail: \email{Karl.Jansen@desy.de}}

\author{Stefan K\"uhn\\
Perimeter Institute for Theoretical Physics, 31 Caroline Street North, Waterloo, ON N2L 2Y5, Canada\\
E-mail: \email{skuhn@perimeterinstitute.ca}}

\abstract{
Tensor Network States are 
ans\"atze for the efficient description of quantum many-body systems. 
Their success for one dimensional problems, together with the fact that they do not suffer from the sign problem and can address 
the simulation of real time evolution, have 
turned them into one of the most promising techniques to study strongly correlated systems.
In the realm of Lattice Gauge Theories they can offer an alternative to standard lattice Monte Carlo calculations, 
which are suited for
static properties and regimes where no sign problem appears.
The application of Tensor Networks to this kind of problems is a young 
but rapidly evolving research field.
This paper reviews some of the recent progress in this area,
and how, using one dimensional models as testbench,
some fundamental milestones have been reached that may pave the way to more ambitious goals.
}

\FullConference{The 36th Annual International Symposium on Lattice Field Theory - LATTICE2018\\
		22-28 July, 2018\\
		Michigan State University, East Lansing, Michigan, USA.}

\begin{document}

\section{Introduction}
\label{sec:intro}

The term Tensor Network States (TNS) encompasses many different families of ans\"atze that can efficiently describe the state of a quantum many-body system. By construction, TNS families correspond to definite entanglement patterns. Quantum information theory provides tools to understand why they can be expected to be good ans\"atze for the physically relevant states, and some of the limitations connected to the simulation algorithms~\cite{Cirac2009rg,Verstraete2008,Orus2014a}.

In the context of numerically investigating strongly correlated systems, TNS have demonstrated to be very powerful, specially for one dimensional problems, with the matrix product state (MPS) ansatz underlying the success of the density matrix renormalization group (DMRG) algorithm~\cite{White1992,Schollwoeck2011,Vidal2003,Verstraete2004}, currently the most precise tool for the study of one-dimensional quantum many-body systems.
The natural generalization of MPS to higher dimensions, Projected Entanglement Pair States (PEPS)~\cite{Verstraete2004b} are good candidates to describe the physics of higher dimensional lattices. Another TNS ansatz, the MERA~\cite{Vidal2007}, can capture critical behavior and has recently been connected to a discrete realization of the AdS/CFT correspondence~\cite{swingle2012}.

Lattice Gauge Theories (LGTs) offer challenging scenarios which can benefit from these techniques. The spatial dimensions and sizes of the systems amenable to TNS studies are still far from those solved by Monte Carlo simulations but, being free from the sign problem, tensor networks can be readily used for problems which more standard techniques cannot easily tackle.
Some early works already combined the gauge invariance with TNS methods~\cite{Byrnes2002,Sugihara2005Z2,Tagliacozzo2011},
and lately there has been a renewed interest on the topic, motivated in part by the prospects of quantum simulation, and also by 
the advances in the understanding and application of tensor networks.

During the last years, different teams have applied TNS algorithms to low dimensional lattice gauge theories
 from different perspectives, and it would be impossible to cover all these efforts in this limited space.
The present discussion is thus restricted to studies in which the LGT is the regularization of a more fundamental quantum 
field theory in the continuum, and the ultimate goal is to solve the latter, much in 
the spirit of lattice QCD. Moreover, we focus on the use of TNS as ansatz for the physical states, in a Hamiltonian formulation 
of the theory. 

Using one dimensional models as a testbed, different 
studies have demonstrated systematically that the MPS ansatz is well suited to describe the relevant physics in LGTs, 
and allows for very precise calculations of low energy and thermal properties, also in scenarios in which 
Monte Carlo techniques are hindered by the sign problem. 
This  article reviews such results, which constitute necessary initial milestones before applying TNS to
more ambitious LGT problems.
The article is structured as follows.
In section~\ref{sec:tns} the main 
TNS ideas and algorithms employed in these studies are discussed.
Section~\ref{sec:models} presents the models that have been used as testbench,
and in section~\ref{sec:results} we review the existing results.
Finally, section \ref{sec:conclu} discusses the status and perspectives of the field. 

\section{Basic concepts}
\label{sec:tns}

\begin{figure}
\begin{center}
\includegraphics[clip, trim=2.5cm 7cm 0.5cm 10cm,width=.75\textwidth]{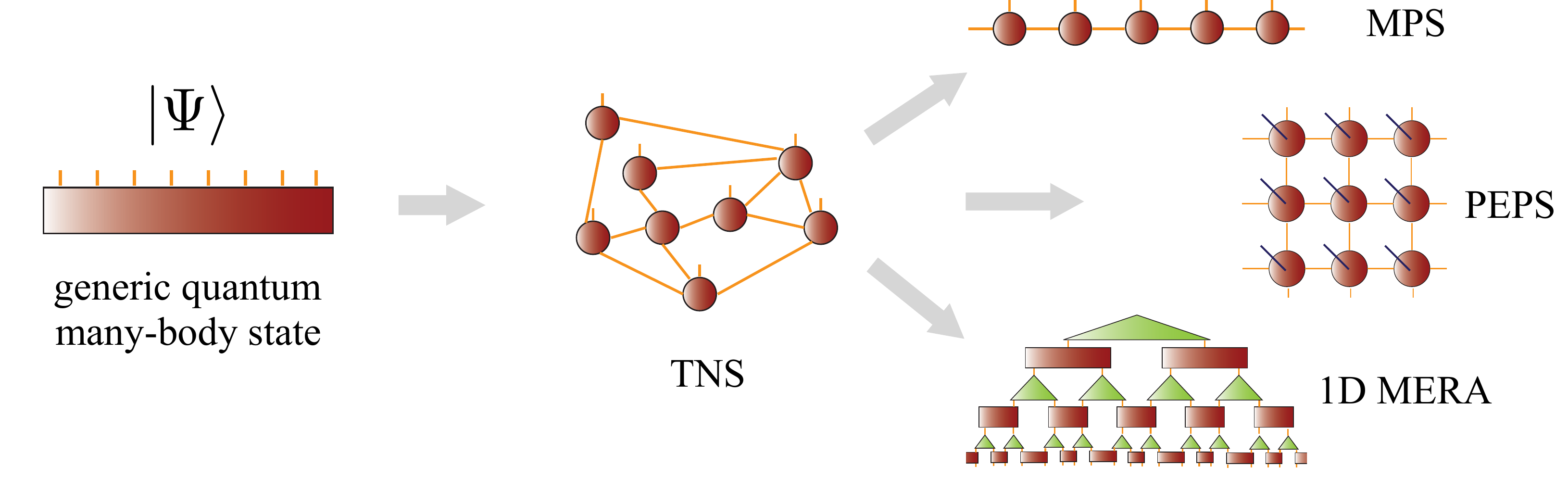}
\vspace{-.25cm}
\caption{Schematic representation of a generic TNS (center), and some of the main TNS families (right).}
\label{fig:TNS}
\end{center}
\vspace{-.5cm}
\end{figure}

If we consider $N$ quantum systems, each one with finite physical dimension $d$,
any quantum many body state can be exactly expressed in a tensor product basis by specifying $d^N$ coefficients, 
$|\Psi\rangle=\sum_{i_1,\ldots i_N=1}^dC_{i_1\ldots i_N} |i_1 \ldots i_N\rangle$.
We can think of the $C_{i_1\ldots i_N}$ coefficients as the $d^N$ components of a 
$N$-legged tensor (which in this context simply denotes a multidimensional array).
The exponential growth of the dimension of the Hilbert space with the number of sites $N$ restricts the
exact calculations to very small system sizes.
However, we may consider restricted ans\"atze where the state coefficients can be expressed as the contraction 
of a certain network of smaller tensors, as illustrated in figure~\ref{fig:TNS}.
Different structures of the network give rise to different TNS families, and naturally correspond to a different 
pattern of entanglement in the represented state.

While random states in the Hilbert space typically have close to maximum entanglement, ground states of physical 
local Hamiltonians behave in a very different manner. They often fulfill what is called the \emph{area law of entanglement}. Namely the
 entropy of a subsystem grows only with the size of the boundary that separates it from the rest. 
In one spatial dimension, the area law has been rigorously proven to hold for ground states of local gapped Hamiltonians,
and for critical cases it is known to be violated only logarithmically~\cite{Hastings2007,Eisert2010}.

This idea justifies the search for adequate families which, in terms of only a polynomial number of parameters and a structure based on the
expected entanglement, are capable
of capturing the essential features of the relevant states.

\subsection{TNS in 1D: Matrix product states}
\label{sec:mps}

The most popular tensor network, for which algorithms are the most efficient, and also the one that is best understood  from 
a mathematical point of view, is the matrix product state family.
A MPS is a state of the form 
$|\Psi\rangle=\sum_{i_1,\ldots i_N=1}^d \mathrm{tr}\left ( A_1^{i_1} A_2^{i_2}\ldots A_N^{i_N} \right ) |i_1 i_2 \ldots i_N\rangle$.
Each $A_k^{j}$ is a matrix of dimensions $D\times D$ (equivalently $A_k$ is a rank-three tensor of dimensions $d\times D\times D$).
In the case of open boundary conditions, the matrices for the first and last sites reduce to $D$-dimensional vectors, and the 
trace can be written as a product of matrices.
The \emph{bond dimension} $D$ determines the
number of parameters in the ansatz, $d N D^2$.
The entanglement between two halves of the chain is upper-bounded by $S(N/2)_{\mathrm{max}}\propto \log D$.
This shows explicitly that MPS satisfy a one-dimensional area law,
which actually makes them expected good ans\"atze for ground states of local Hamiltonians.
A single tensor can be used to parametrize a translationally invariant state, where $A_k=A$  for every site $k$, either for 
a finite periodic chain, or for an infinite system. With this ansatz, named uniform MPS (or uMPS),
it is possible to directly work in the thermodynamic limit  (see e.g.~\cite{Verstraete2008,Schollwoeck2011} for reviews).

Global and local symmetries of the state can be encoded in properties of the individual tensors \cite{Tagliacozzo2014,Haegeman2015,Zohar2015b}.
Investigating which classes of states can be represented by TNS with different symmetry restrictions is an active
field of research.
From the point of view of numerical calculations, explicitly imposing the gauge symmetry at the level of the individual tensors
can reduce the number of variational parameters (and thus the computational cost), ensuring at the same time
that the ansatz is in the desired sector, for instance, the physical gauge-invariant subspace in the case of a lattice gauge theory.

Different numerical methods exist to solve a given physical problem with MPS.
The following paragraphs briefly review the main aspects of the ones that have been so far applied to LGT problems.
Many of these algorithms have been generalized to other TNS families, although 
in general this involves much higher computational costs.

\subsection{Variational optimization}

In order to find a MPS approximation to the ground state of a local Hamiltonian,
it is possible to variationally minimize the energy within the MPS family with fixed bond dimension 
by sequentially optimizing the tensors in the ansatz one at a time, and iterating the 
procedure until convergence.
This variational strategy basically describes the extremely successful DMRG algorithm
and can easily be extended to obtain low lying excitations by searching for the lowest energy state that is 
orthogonal to previously computed levels~\cite{Verstraete2008,Schollwoeck2011}.

In the case of uMPS in the thermodynamic limit, it is also possible to parametrize the excitations as a single (or few)
 different tensors over the uniform background that corresponds to the ground state. 
A state of well-defined momentum can be constructed as an appropriate superposition of all possible translations of 
this construction, and the tensor parametrizing the excitation can also be optimized variationally~\cite{Haegeman2011,Haegeman2013post}.

\subsection{Time evolution}

It is also possible to find an approximation to the ground state using imaginary (or Euclidean) time evolution.
The most widely used strategy to evolve a MPS, in real or imaginary time, is  based on
a Suzuki-Trotter expansion of the exponential evolution operator~\cite{Verstraete2008,Schollwoeck2011}. The time is discretized in small steps, whose corresponding 
evolution operators can be expressed as products of local terms (e.g. two-body operators in the case of 
a nearest-neighbor model). The action of each such term on the MPS yields in general a MPS of larger bond dimension,
so the result needs to be truncated, a step that may introduce a numerical error.
This strategy is in practice limited to local Hamiltonians, but more sophisticated algorithms exist that can also cope
with long range interactions \cite{Haegeman2011}.

The algorithm is essentially the same for real or imaginary time evolution, but in the former problem the entanglement in the physical state may grow very fast (up to linearly with the elapsed time), which implies that the bond dimension required to keep an accurate description of the time-dependent state as a MPS will increase exponentially. 
This in practice limits the feasibility of real time simulations to short times or close to equilibrium situations.

\subsection{Thermal states}

Quantum many-body operators can be expressed in a tensor product basis
in which every element is the tensor product of  one element of a local basis for each site of the system.
If the coefficients of the expansion have a MPS structure, we talk about a matrix product operator (MPO), 
an extremely useful form of writing Hamiltonians and approximating evolution operators.
Thermal equilibrium and other mixed states are described by a density operator,
and the MPO ansatz can be applied to approximate them  \cite{Verstraete2004a,Zwolak2004,Pirvu2010a}.
Moreover, it is known to do it efficiently 
 in the case of local Hamiltonians (also in higher dimensions)~\cite{Hastings2006}.
 
The positivity constraint, necessary to ensure that the MPO ansatz encodes a physical state,
is not easy to impose at the level of the individual tensors.
One way to overcome this problem is to use a more restricted MPO ansatz, referred to as purification,
in which each of the tensors has a positive structure~\cite{Verstraete2004a,delasCuevas2013puri}.
This is equivalent to assuming a certain MPS form for a pure state that describes an enlarged system,
and in the case of the thermal equilibrium states corresponds to a thermofield doubling.
The purified state for inverse temperature $\beta$ is then, up to normalization, the result of applying the 
exponential operator $e^{-\beta H/2}$ to a maximally entangled state of system and ancilla.
With the appropriate ordering of system and ancilla sites, such state is a trivial MPS, and
the action of the exponential can be approximated using an algorithm for imaginary time evolution.
This yields a very stable and widely used algorithm to find thermal states with TNS.
 
\section{Models}
\label{sec:models}

One dimensional models are a natural testbench to investigate the performance of TNS numerical methods.
The studies reported here have considered several models of fermions interacting with gauge fields. In the discrete formulation,
fermions reside on the vertices of the lattice
and gauge degrees of freedom on the links between them
and the MPS ansatz can be formulated in a basis which is the tensor product of the local
basis for fermionic and gauge degrees of freedom.

\paragraph{Schwinger model.}
One of the key example models is the Schwinger model, or QED in one spatial dimension.
Arguably the simplest gauge theory with matter, 
it exhibits phenomena, as confinement and chiral symmetry breaking, common to more complex gauge theories,
and a phase transition in presence of a 
background field (which is the equivalent in this case of a topological theta term).
In the temporal gauge, and using Kogut-Susskind formulation with staggered fermions,
the discretized Hamiltonian in the absence of background field
reads~\cite{Kogut1975},
\begin{equation}
H=-\frac{i}{2 a}\sum_n \left (  \phi^{\dagger}_n e^{i\theta_n}  \phi_{n+1} - \mathrm{h.c.}\right ) 
+m\sum_n(-1)^n\phi_n^{\dagger}\phi_n
+\frac{a g^2}{2} \sum_n L_n^2,
\label{eq:schwinger}
\end{equation}
where $ \phi_{n}^{\dagger}$
represents the creation operator of a spinless fermion on lattice site $n$, and
$U_n=e^{i\theta_n}$ is the link operator between sites $n$ and $n+1$.
$L_n$, canonical conjugate to $\theta_n$, corresponds to the electric field on the link.
Gauss law appears as an additional constraint that has to be satisfied
by physical states, namely
$L_n-L_{n-1}=\phi_n^{\dagger}\phi_n-\frac{1}{2}\left [1-(-1)^n\right]$.
The parameters of the model are the fermion mass $m$ and the coupling $g$ (the only parameters in the continuum), and the 
 lattice spacing $a$.
It is convenient to rescale the Hamiltonian and work with adimensional parameters, $x=1/(a g)^2$, $\mu=2\sqrt{x}m/g$, 
such that the continuum limit corresponds to $x\to \infty$.
 The local Hilbert space basis for the fermionic sites can be labeled by the occupation of the mode, $n_m\in\{0,1\}$ (for site $m$),
while the basis elements for the links can be labeled by the integer eigenvalues of $L_m$, $\ell_m\in \mathbb{Z}$.

\paragraph{Multiflavor Schwinger model.} 
A generalization of the model above consists in considering several different fermion flavors,
each one with a different chemical potential $\kappa_f$.
Except for certain specific choices of the parameters, conventional Monte Carlo techniques will encounter here a sign problem.
The modified Hamiltonian can be written
\begin{equation}
H = -\frac{i}{2a}\sum_{n=0}^{N-2}\sum_{f=0}^{F-1}\left(\phi^\dagger_{n,f}e^{i\theta_n}\phi_{n+1,f}-\mathrm{h.c.}\right)
+\sum_{n=0}^{N-1}\sum_{f=0}^{F-1}\left(m_f(-1)^n +\kappa_f \right)\phi^\dagger_{n,f}\phi_{n,f}
+ \frac{ag^2}{2}\sum_{n=0}^{N-2} L_n^2,
\label{eq:schwinger-multi}
\end{equation}
which can also be expressed in terms of adimensional parameters $x=1/(a g)^2$, $\mu_f=2\sqrt{x}m/g$,  
$\nu_f=2\sqrt{x}\kappa_f/g$. 
Gauss law now reads
$L_n - L_{n-1} = \sum_{f=0}^{F-1}\left[ \phi^\dagger_{n,f}\phi_{n,f}-\frac{1}{2}(1-(-1)^n)\right]$,
involving a sum over all fermion flavors.
While the structure of the links does not change with respect to the previous case, the local Hilbert space for the fermionic site $i$ is described by $F$ occupation numbers corresponding to each flavor,
$\otimes_f |n_{m,f}\rangle$.
The structure of the ground state in this case shows a series of transitions between phases with different imbalances~\cite{Lohmayer2013}.
Thus, it is an ideal problem to probe the performance of TNS in a simple scenario  
that yet is difficult for standard Monte Carlo methods.

\paragraph{SU(2) in 1+1 D.} Another fundamental ingredient that a numerical technique has to work with 
is a non-Abelian gauge symmetry.
The simplest model we can consider is a 1+1 dimensional SU(2) theory,
for which the discrete Hamiltonian in the staggered fermion formulation reads
\begin{equation}
H=\frac{1}{2a} \sum_{n=1}^{N-1}\sum_{\ell,\ell'=1}^2 \left({\phi_{n}^{\ell}}^\dagger U_{n}^{\ell\ell'}\phi_{n+1}^{\ell'}+\mathrm{h.c.}\right)
+ m\sum_{n=1}^N\sum_{\ell=1}^2 (-1)^n{\phi_{n}^{\ell}}^\dagger\phi_{n}^{\ell} +\frac{ag^2}{2}\sum_{n=1}^{N-1} \mathbf{J}_{n}^2.
\label{eq:su2}
\end{equation}
Now the link operators $U^{\ell\ell'}_n$ are SU(2) matrices in the fundamental representation, and can be interpreted as rotation matrices.
Gauss law is also a non-Abelian constraint, 
$G^\tau_m|\Psi\rangle=0$, $\forall m, \tau$,
with generators $G^\tau_m =L^\tau_m-R^\tau_{m-1}-Q^\tau_m$, where 
$Q_m^\tau=\sum_{\ell=1}^2\frac{1}{2}{\phi_m^\ell}^\dagger\sigma^\tau_{\ell \ell'}\phi_m^{\ell''}$ are the components of the non-Abelian charge 
at site $m$.\footnote{If present, external charges also need to be included.}
 The generators of left and right gauge transformations on the link $L^\tau$ and $R^\tau$, $\tau\in\{x,y,z\}$, can be interpreted as color-electric fields,
 and
 the color-electric flux energy is $\mathbf{J}_{m}^2 = \sum_\tau L^\tau_mL^\tau_m = \sum_\tau R^\tau_mR^\tau_m$.
 Now the Hilbert space of the $m$-th link is analogous to that of a quantum rigid rotator with total angular momentum $\mathbf{J_{m}^2}$, with 
  body-fixed angular momentum operators $L^{\tau}_m$ and space-fixed $R^{\tau}_m$,
  and the basis elements can, thus, be labeled by the eigenvalues of $\mathbf{J}_{m}$, $L^{z}_m$ and $R^{z}_m$, 
as $ |j_m \ell_m \ell'_m\rangle$.
For the fermions, the Fock basis for each site now specifies the occupation numbers of each of the two color components, 
$|n_m^1,n_m^2\rangle$.

\section{Numerical LGT calculations with TNS}
\label{sec:results}

The study of the models presented in section~\ref{sec:models} with MPS techniques requires some previous considerations.
\begin{itemize}
\item
The MPS formulation presented in Section~\ref{sec:mps} requires a finite dimensional Hilbert space for every site of the system.
Although  the gauge degrees of freedom do not fulfill this requirement,
 it is possible to truncate the basis of each link to a finite size.
Numerical evidence suggests that at least for low energy states, a small truncation will be enough, 
even close to the continuum limit
\cite{Buyens2017}.
\item
TNS can work with fermionic degrees of freedom without running into a sign problem, or
increasing the computational complexity of the algorithms~\cite{Kraus2010,Corboz2010,Pineda2010}.
But the anticommutation of the corresponding operators requires specific 
attention in each step of the algorithm.
In the one dimensional case, this can be simplified
by applying a Jordan-Wigner transformation that maps the fermions to spins,
which then can be treated by the most standard MPS algorithms.
This strategy si adopted by the works we discuss next.
\item
Only MPS that satisfy Gauss law are acceptable ans\"atze for  
physical states. The constraint can be imposed as a local symmetry at the level of the individual 
tensors \cite{Buyens2013,Rico2013}, such that only gauge invariant TNS are included in the ansatz.
This can be generalized to higher dimensions~\cite{Tagliacozzo2014,Zohar2015,Haegeman2015,zohar2018mc}.
However, in the one dimensional case there is also the possibility to completely integrate out the gauge degrees of freedom, 
as they are completely fixed by the fermionic content, and to
formulate the problem in the strictly physical space, thus dealing with the problem in a most efficient 
way in terms of number of variables, but at the cost of introducing non-local terms in the Hamiltonian.
Notice that in this case, no truncation of the gauge degrees of freedom is needed, in principle~\cite{Banuls2013,Banuls2017}.
\end{itemize}

Over the last few years, different numerical studies have approached the problems above with various choices
for boundary conditions and symmetries, and have systematically analyzed the continuum limits and the errors involved.
As a consequence, the abundance of nowadays available results  
demonstrates beyond doubt the feasibility of the MPS ansatz for the physical states of gauge theories.
In the following subsections, we review concrete numerical results that illustrate
some of the most significant steps in this program.

\subsection{Spectral properties}

\subsubsection{Mass gaps}

\begin{figure}
\vspace{-.3cm}
\begin{center}
\includegraphics[width=.45\textwidth]{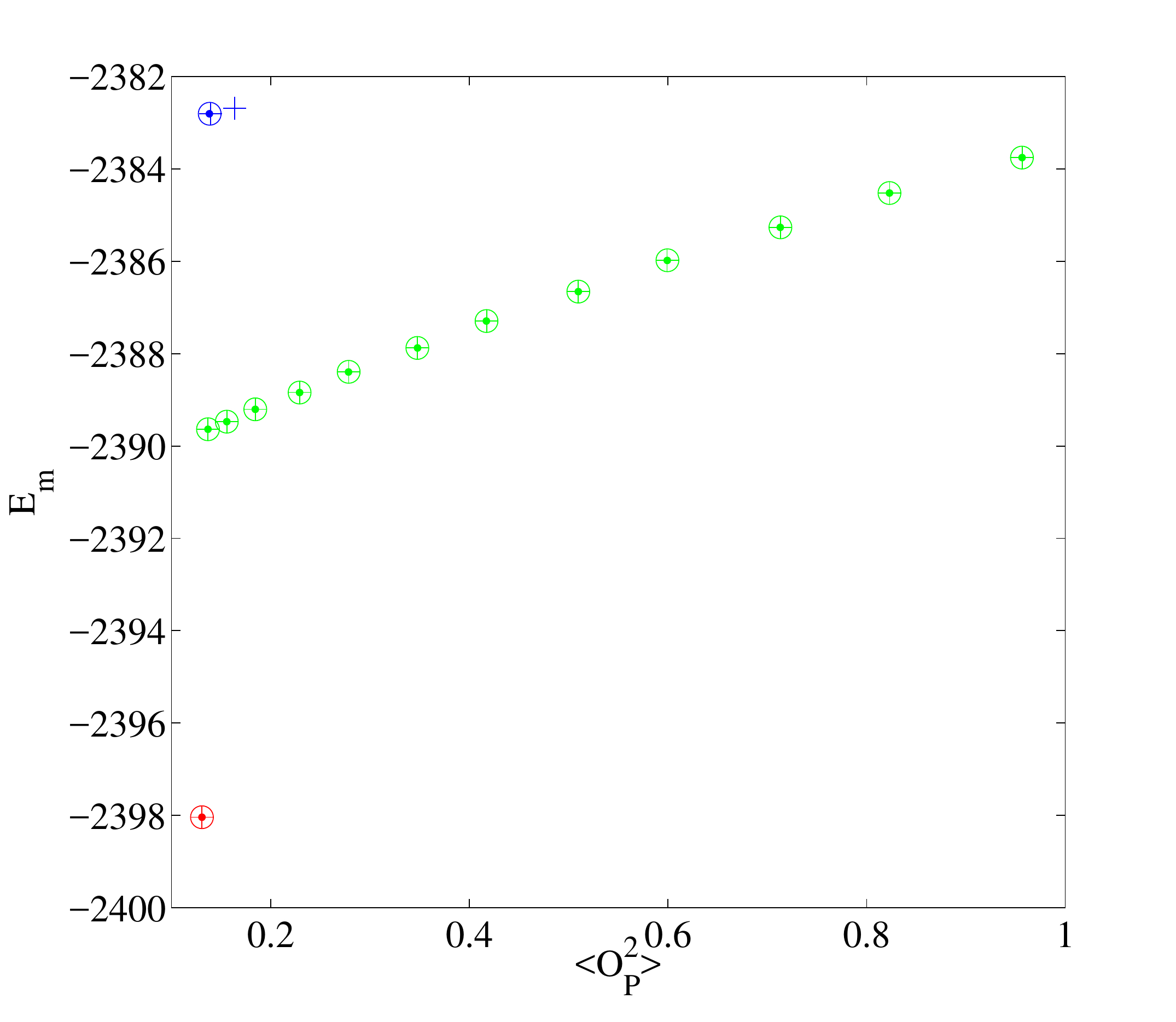}
\includegraphics[width=.4\textwidth]{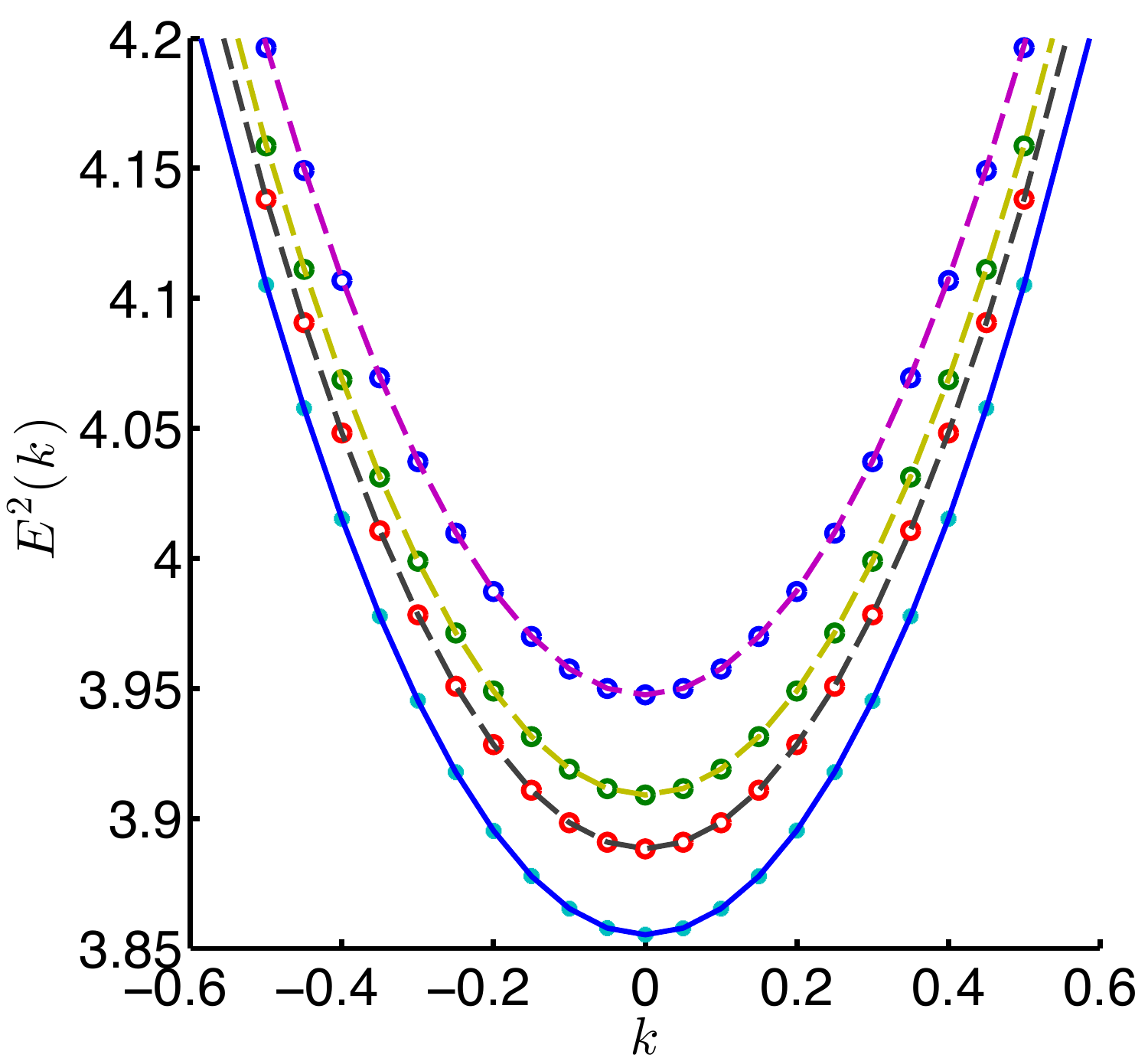}\\
\caption{
Dispersion relations for the massive Schwinger model. The left panel shows the results for open boundary MPS
in the physical subspace~\cite{Banuls2013},
for fermion mass $m/g=0.125$, 
system size $N=160$ and lattice spacing 
given by $x=1/(ga)^2=25$.
The colors indicate the ground state (red), vector (green) and scalar (blue) branches, and symbols correspond to bond dimensions $D=40$ (crosses), $80$ (circles), and $100$ (dots).
The right plot~\cite{Buyens2013}
 shows the vector excitation energies for non-zero momenta, computed 
using the uMPS ansatz and already converged in bond dimension, 
for $m/g = 0.75$, and lattice spacings $x = 100, 300, 800$ (circles), as well as the continuum estimates (stars).
The curves fit the data to the relativistic dispersion relation $E^2 = k^2 + M^2(x)$. 
\emph{Right figure reproduced with permission from~\cite{Buyens2013}, @2014 APS.}
}
\label{fig:dispersionSchw}
\end{center}
\vspace{-.45cm}
\end{figure}

The MPS ansatz can be used to approximate the lowest energy eigenstates of the lattice model, and
after taking the appropriate continuum limit, calculate the particle content of the continuum theory.
In the case of the Schwinger model, the results can be benchmarked against the
analytical solution at zero fermion mass, to assess the performance of the TNS calculations.
The first impressive result was obtained early on in~\cite{Byrnes2002} using the original DMRG algorithm 
and, compared to previous numerical results, improved by several orders of magnitude the precision of the 
ground state energy and first (vector) mass gap in the continuum for massive and massless fermions. 
But only the TNS formulation allowed the power of DMRG to be extended for higher excitations, 
time evolution and other problems.
More recently, in~\cite{Banuls2013}, using finite MPS with open boundary conditions in the physical space, and in \cite{Buyens2013}, 
using a translationally and gauge invariant uMPS, calculations of the vector and scalar 
mass gaps were performed with unprecedented precision. 
In \cite{Buyens2013}, the second vector excitation was accessed numerically for the first time.

Each MPS simulation is run for a finite lattice spacing and with a certain maximum bond dimension. In the case of finite systems
as in ~\cite{Banuls2013}, also for a fixed system size. For each such case, the algorithm
identifies the ground state and several excitations, which differ in quantum numbers (e.g. left panel of figure~\ref{fig:dispersionSchw}).
To reach the continuum limit, several extrapolation steps are necessary.
\begin{enumerate}
\item
Bond dimension $D$: increasing the size of the tensors in the simulation improves the precision 
systematically. 
In the case of the Schwinger model, low energy states indeed converge very fast in $D$, as illustrated by the left
plot in figure~\ref{fig:dispersionSchw}, where data for $D=40$, $80$, and $100$ collapse on top 
of each other for a system of $N=160$ sites.\footnote{These numbers are to be compared to the largest possible bond dimension for a system of this size, $D_{\mathrm{max}}=2^{80}$.}
\item
System size $N\to \infty$. This is necessary only in case of finite system simulations, which have in exchange the 
advantage of being unbiased, and able to cope with non-translationally-invariant scenarios. 
The uMPS works directly in the thermodynamic limit and does not require this step, but it assumes a local form for the excitations, 
and may find worse convergence if they are extended, as is the case for very small fermion mass.
\item
Lattice spacing $x\to \infty$. 
To approach the continuum limit for a fixed fermion mass, simulations with large enough values of $x$ 
(and the corresponding adimensional masses $\mu$) are required. 
In the continuum limit, the theory becomes critical, and in principle harder for the MPS algorithms.
Yet, the large values of $x$ (correspondingly small lattice spacings) reachable with MPS, as compared to Monte Carlo calculations,
allow for more precise determination of the continuum values than with any other numerical technique, in 
the general case.\footnote{For the massless fermion case, the most precise results have been obtained
using strong coupling expansion~\cite{Cichy2013}.}
It is actually possible to discriminate the presence of cubic or other higher order corrections in the lattice spacing.

\end{enumerate}

The analysis shows that the main source of error is the continuum limit, a systematic uncertainty that can,
nevertheless, be taken into account by a standard error analysis procedure, similar to those  in 
data analysis for Lattice QCD~\cite{Banuls2013}.
The extremely precise estimations of the mass spectrum obtained with  different MPS algorithms
and for different fermion masses (far from the perturbative regime)
are shown in table~\ref{tab:specSchw}.

\begin{table}
\begin{center}
\begin{tabular}{|c|c|c|c|c|}
\hline
&\multicolumn{2}{c|}{$M_V/g$}
&\multicolumn{2}{c|}{$M_S/g$} 
\\ 
\cline{2-5}
$m/g$ &
OBC \cite{Banuls2013} &
uMPS \cite{Buyens2013} &
OBC \cite{Banuls2013} &
uMPS \cite{Buyens2013} \\
\hline
0	&
0.56421(9) &
0.56418(2)& 
1.1283(10) &
- \\
\hline
0.125 &
0.53953(5) &
0.539491(8)&
1.2155(28)  &  1.222(4) \\ \hline
0.25 &
0.51922(5) & 0.51917(2)
&
1.2239(22)  &  1.2282(4)\\ \hline
0.5 &
0.48749(3) & 0.487473(7)
&
1.1998(17)  &  1.2004(1)\\ \hline
\end{tabular}
\caption{Binding energies of the vector and scalar particles and their reported errors obtained with  open boundary finite MPS with gauge degrees of freedom integrated out (left columns) or gauge invariant uniform MPS (right column) simulations. For the massless fermion, the results can be compared to the analytical values 
for the vector and scalar binding energies, respectively $M_V/g=0.5641895$
and $M_S/g=1.12838$.}
\label{tab:specSchw}
\end{center}
\vspace{-.5cm}
\end{table}

A similar analysis has also been performed for the 1+1 dimensional SU(2) Hamiltonian \cite{Banuls2017}.
In that case, the continuum theory becomes massless as $m/g\to 0$, and 
the MPS simulations allow a study of critical exponents.

\subsubsection{Chiral condensate}

From the MPS optimized for a certain state of the spectrum, 
e.g.\ the vacuum, any local or few body observable can be efficiently computed. 
This allows the study of different quantities in the continuum limit.
One such study has been performed for the fermionic condensate,
$\langle \overline{\Psi}(x) \Psi(x) \rangle/g$, the order parameter of chiral symmetry breaking.
This quantity is UV divergent for massive fermions, the divergence  
coming solely from the free ($g=0$) theory.
The algorithms mentioned in the previous section~\cite{Banuls2013a,Buyens2014}
also explored this observable in the massless and massive regimes.
The precision of the results allowed for extracting the divergent behavior and identifying the leading 
corrections with the lattice spacing, and produced the most precise estimates of the chiral condensate in the
ground state for non-vanishing fermion masses.
Similar results have also been obtained with the related iDMRG algorithm~\cite{Zapp2017}.

\subsubsection{The entropy of the vacuum}

\begin{figure}
\begin{center}
\includegraphics[width=.35\textwidth]{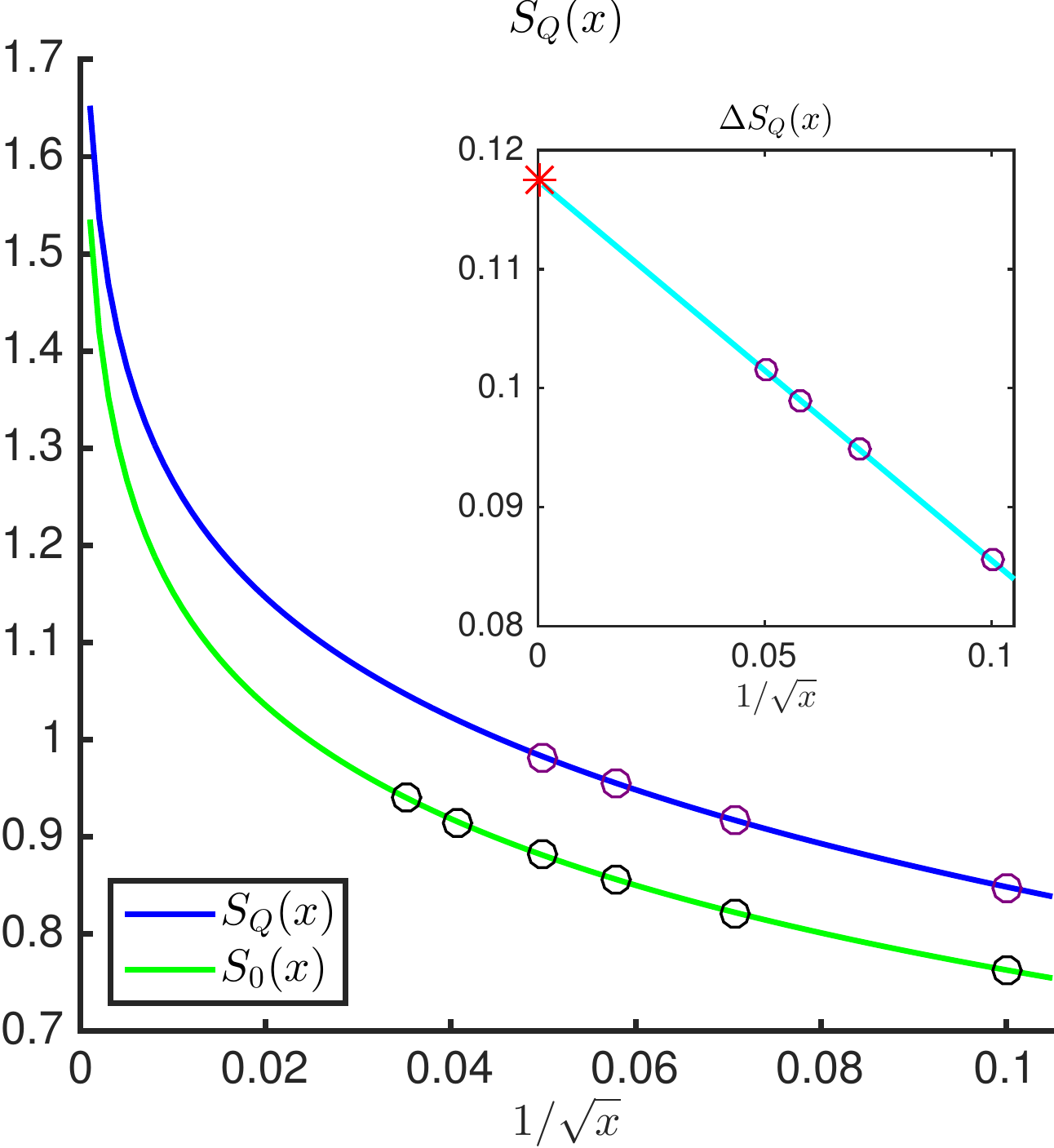}
\hspace{.5cm}
\includegraphics[width=.5\textwidth]{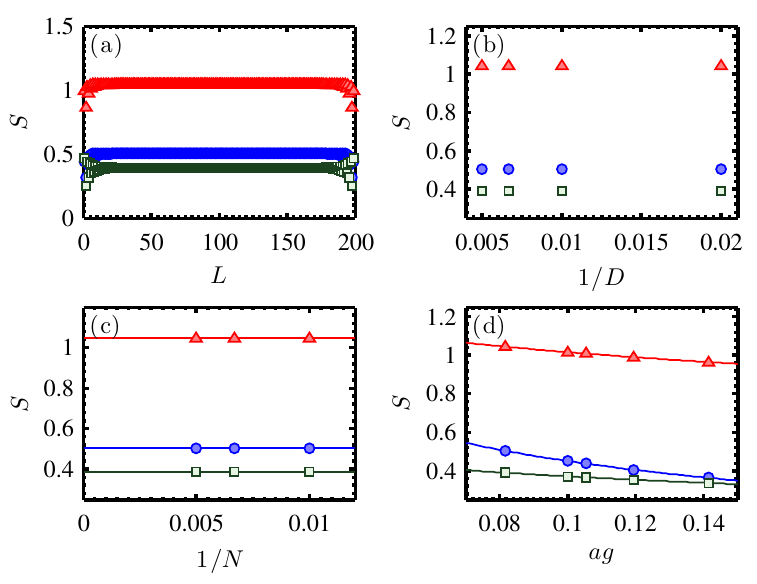}
\vspace{-.2cm}
\caption{Entanglement entropy of the half chain in the vacuum of a lattice gauge theory. The leftmost panel
shows the extrapolation to the continuum in the case of
 the Schwinger model, for $m/g=0.25$~\cite{Buyens2015}.
 The four plots on the right correspond to the SU(2) model~\cite{Banuls2017} and show, for $m/g=0.8$, the different 
 contributions to the entanglement entropy for each possible cut of the chain (a), as well as the bond dimension (b), 
 system size (c) and continuum (d) extrapolations. 
 \emph{Left figure reproduced with permission from~\cite{Buyens2015}, @2016 APS.}
 }
\label{fig:entropydiv}
\end{center}
\vspace{-.45cm}
\end{figure}

Of particular interest is the numerical analysis of the bipartite entanglement entropy of the vacuum,
a quantity that can be accessed very efficiently in MPS.
On the one hand, in the continuum limit, which corresponds to a massive relativistic QFT,
the entanglement entropy is predicted to diverge 
as $S=(c/6)\log_2(\xi/a)$ \cite{Calabrese2004},
where $c$  is the central charge of the conformal field
theory describing the system at the critical point.
The MPS data accurately capture this behavior, as has been explicitly checked
for the Schwinger~\cite{Buyens2015} and the SU(2)~\cite{Banuls2017} models (see figure \ref{fig:entropydiv}),
where the numerical data were fitted to the logarithmic divergence, and the value of $c$ was extracted.
In ~\cite{Buyens2015}, this allowed the definition of a renormalized entropy difference
in the presence of external charges.
On the other hand, the definition of the entanglement entropy of a gauge invariant lattice system
entails some subtleties, since the gauge constraints are not local with respect to 
a straightforward bipartition of the Hilbert space~\cite{Ghosh2015,VanAcoleyen2016}. 
Instead, different contributions to the entropy can be identified, of which only one is distillable, and the others
arise from the gauge invariance of the state.
With the MPS ansatz, it is possible to compute these different contributions (see right part of figure \ref{fig:entropydiv}) and 
to study their scaling and dependence on the parameters of the model.

\subsection{Thermal properties}

\begin{figure}
\vspace{-.5cm}
\begin{center}
\includegraphics[width=.35\textwidth]{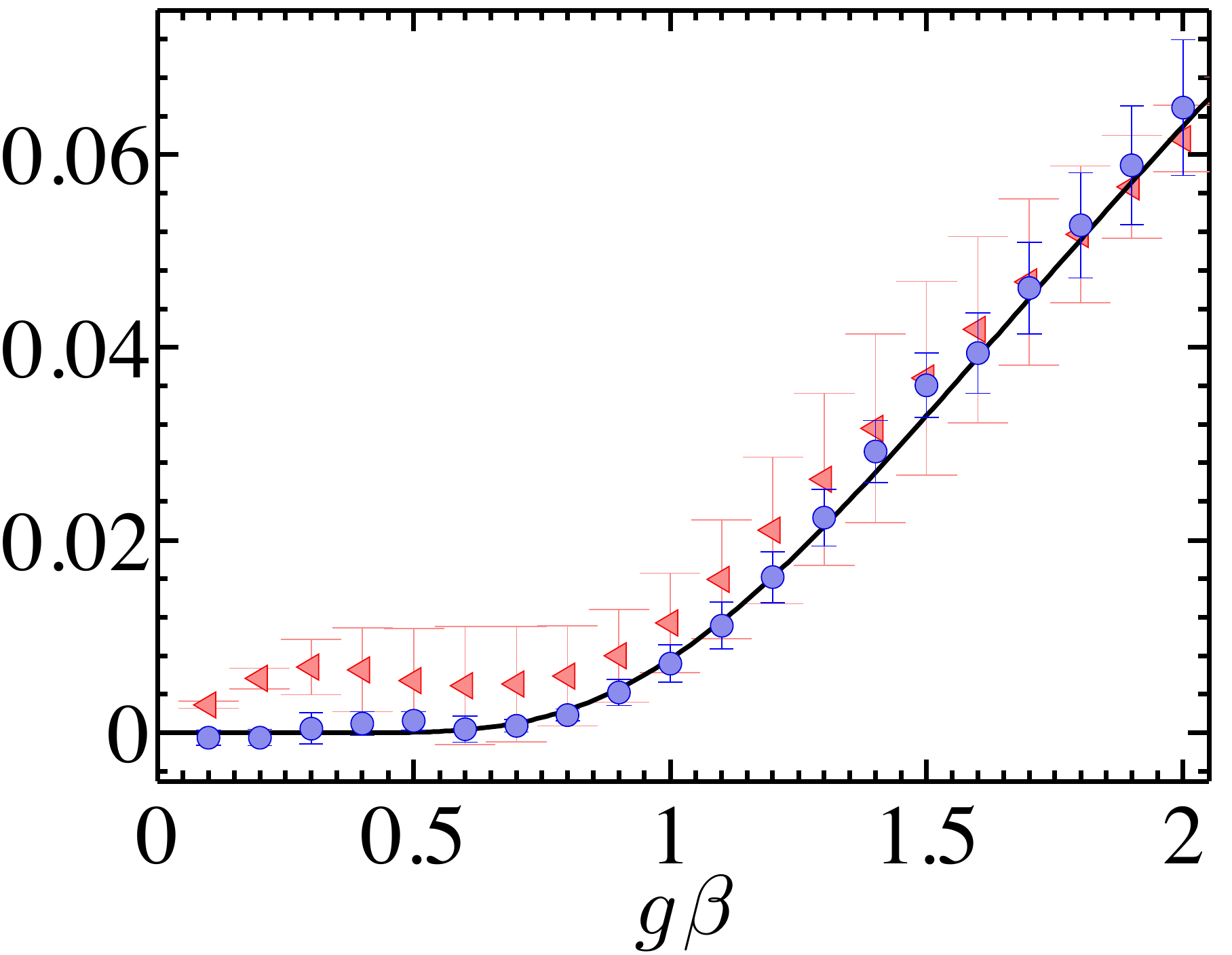}
\hspace{.15\textwidth}
\includegraphics[width=.32\textwidth]{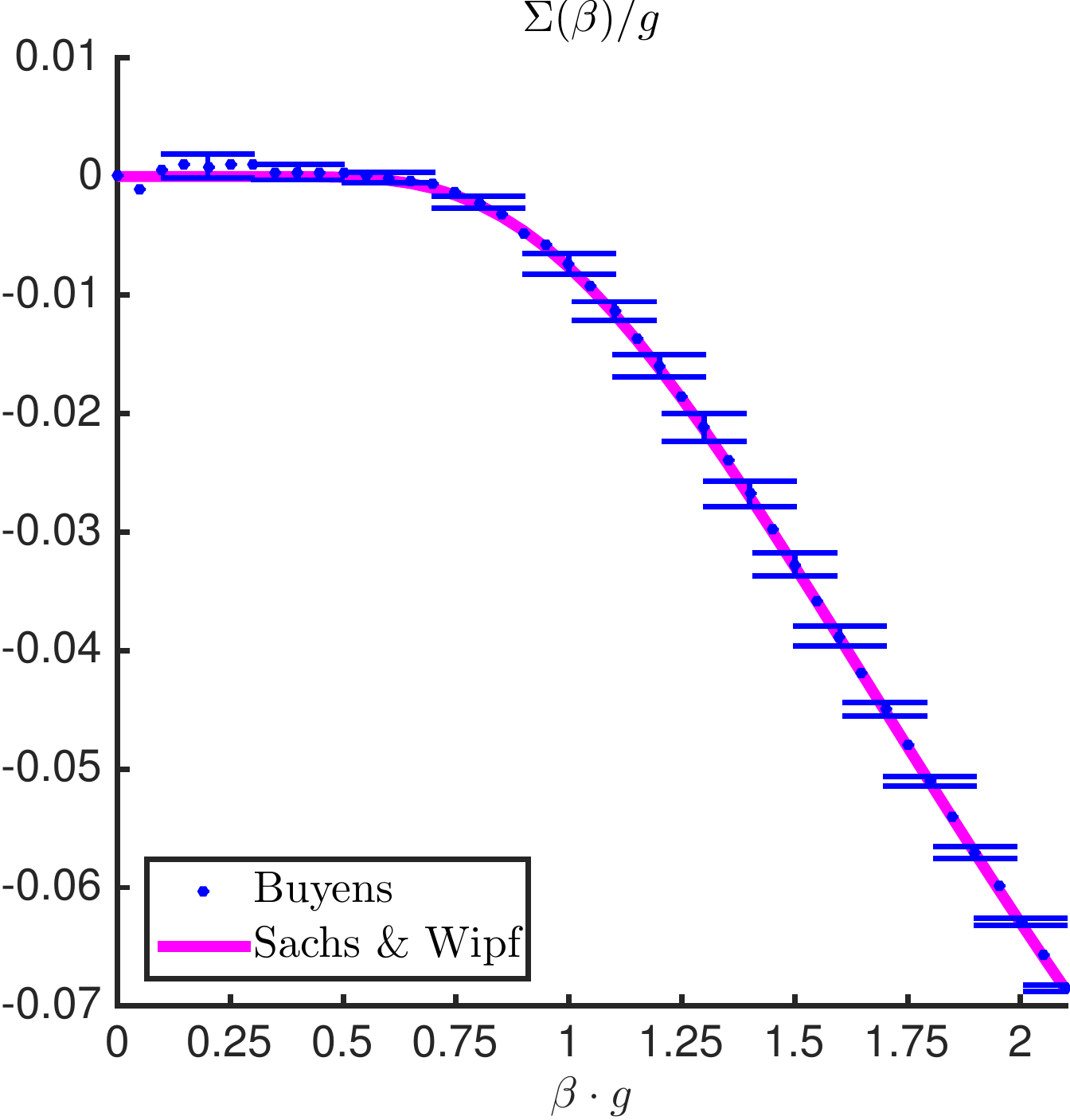}
\caption{Temperature dependence of the chiral condensate in the continuum 
for massless fermions, from finite size MPS on the physical space~\cite{Banuls2015} (left) and gauge invariant uMPS (right, reproduced from Ref.~\cite{Buyens2016}).
Notice both studies used an opposite change convention.
In both cases, the analytical results for the restoration of chiral symmetry (solid line) are recovered with good accuracy.
On the left plot, two sets of data are compared, obtained using different approximations for the imaginary time evolution operator.} 
\label{fig:thermal1}
\end{center}
\vspace{-.4cm}
\end{figure}

Using the MPO ansatz for the density operator it is possible to study temperature dependent properties of LGTs. 
This has been applied to extract the thermal evolution of the chiral condensate in 
the massless and massive Schwinger model~\cite{Banuls2015,Buyens2016}.
Analytical calculations for $m/g=0$ predict a smooth restoration of the chiral symmetry at high 
temperatures, 
 a feature that the MPS calculations reproduce with good 
fidelity (figure~\ref{fig:thermal1}).

The approximation of thermal equilibrium states is usually done by evolving the thermofield double from infinite 
temperature to the desired $\beta=1/T$. This introduces an additional parameter, namely the width of the imaginary time 
step, which has to be taken into account in the extrapolations towards the continuum. When using Suzuki-Trotter 
approximation this is nevertheless a well-controlled error and, as in the mass gap calculations, the 
largest uncertainty source is found to be the continuum extrapolation.
This explains why the quality of the numerical results 
worsens towards infinite temperature: while the MPO approximation for the thermal state 
of a given spin chain is actually easier at high temperature, much smaller lattice spacings are 
needed to correctly capture the lattice effects in that regime.

MPS allow the same calculations also for massive fermions and over the whole range of temperatures,
reaching regimes for which no previous precise numerics existed~\cite{Banuls2016,Buyens2016}.

\subsection{Finite density}

\begin{figure}
\vspace{-.1cm}
\begin{center}
\includegraphics[width=.43\textwidth]{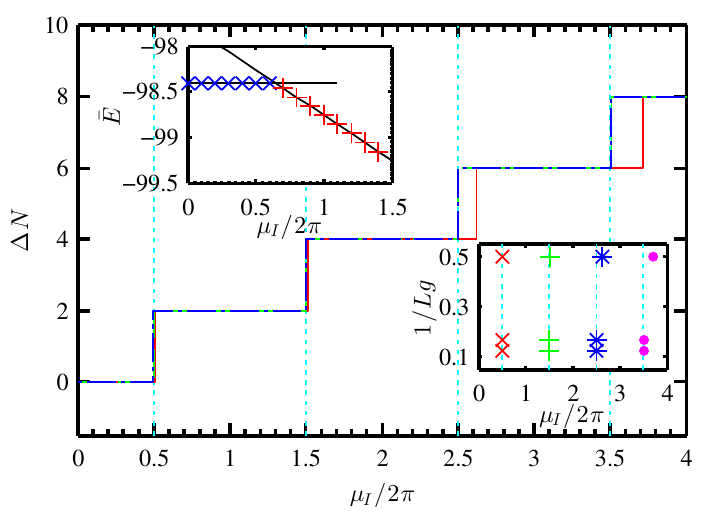}
\hspace{.05\textwidth}
\includegraphics[width=.43\textwidth]{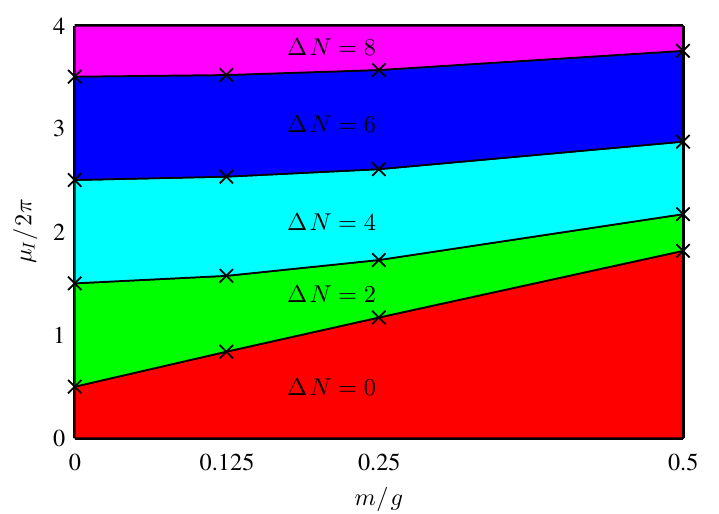}
\caption{Phase diagram of the multiflavor Schwinger model, from \cite{Banuls2016a}. 
The left plot shows the continuum extrapolation of the isospin number for the case of massless fermions, 
as a function of the (normalized) difference of chemical potentials and 
for different physical volumes. The steps show the location of the transitions, which can be compared to the analytical 
prediction (vertical lines).
The right plot shows the phase diagram in the plane chemical potential difference vs. fermion mass,
for fixed volume $Lg=8$,
with regions of different colors representing phases characterized by
different isospin number.}
\label{fig:multifl}
\end{center}
\vspace{-.4cm}
\end{figure}

One of the most prominent features of TNS techniques is that they are free from the sign problem. 
The first explicit demonstration of this capability for a lattice calculation including the continuum limit extrapolation
has been performed in~\cite{Banuls2016a} for the Schwinger model with two fermion flavors and non-vanishing 
chemical potential defined in Eq.\ (\ref{eq:schwinger-multi}).

The case of massless fermions can be solved analytically in the continuum, and at fixed volume
exhibits a rich structure of phases, characterized by the \emph{isospin number}  $\Delta N$ (imbalance between fermion number for both flavors) and separated by 
first order phase transitions. 
These features are recovered from the MPS calculations carried out at constant finite physical volume,
as shown in the left panel of figure~\ref{fig:multifl}.
By performing the same analysis for massive fermions (right panel of the figure),
the phase diagram can be explored beyond the Monte Carlo accessible regimes.

\subsection{Real time evolution}

Another interesting aspect of TNS methods is the possibility to simulate out of equilibrium dynamics.
In the case of the massive Schwinger model, the first real time simulations have already been performed 
with MPS~\cite{Buyens2013,Buyens2016b}.
Although these studies did not include a full continuum extrapolation as the ones 
above, the simulations were 
carried out close to the continuum (at very small lattice spacing, $x=100$).
The results allowed a study of the Schwinger pair production mechanism for quenches of varying 
strength, well beyond the linear response regime and, for moderate times,
found effects that escape other approximate real-time methods for the lattice.

\section{Discussion and perspectives}
\label{sec:conclu}
\vspace{-.1cm}

This paper has focused on
 studies where the TNS techniques are seen
as a numerical tool, alternative to standard LGT methods,
 for solving continuum quantum field theories.
The excellent results achieved for one dimensional problems have demonstrated 
beyond doubt the validity of the ansatz for the states of the theory, and its ability 
to capture all the relevant physics, also in scenarios 
that present difficulties for standard methods.
A more ambitious goal is to extend these simulations to higher dimensional TNS,  as PEPS.
Such generalization is not trivial, but 
 the recent algorithmic advances~\cite{Corboz2016a,Vanderstraeten2016a},
 the development of gauge invariant formulations
 \cite{Tagliacozzo2014,Haegeman2015,Zohar2015,zohar2018mc}, 
 and the success of the one dimensional
calculations constitute a solid motivation
to pursue this endeavor.

As advanced in the introduction, there are other research directions connecting TNS and LGT.
First of all, it is also possible to use a tensor network to represent the path integral for a lattice gauge model in space-time,
and apply TNS algorithms 
to approximate the contraction without suffering from a sign problem, even for two spatial dimensions 
(see e.g.~\cite{Meurice2013,Liu2013,Shimizu2014,Kawauchi2016}). 
Another active area of research is using TNS
 to study LGT that do not necessarily have a continuum limit,
such as quantum link models~\cite{Chandrasekharan1997}.
Successful simulations have been performed for one-dimensional U(1) and SU(2) models
for static and dynamical quantities~\cite{Rico2013,Pichler2015,Silvi2016}.
A  topic closely related to this is the possible quantum simulation of gauge theories in the lab
using ultracold atoms~\cite{Zohar2015a}.
Actually a first pioneering experiment has already been performed for the quantum simulation of the Schwinger model
using trapped ions~\cite{Martinez2016}.
The TNS requirements mentioned above, namely a Hamiltonian formulation, and the finite-dimensional
Hilbert space of local degrees of freedom, are also necessary for a quantum simulation with such systems,
and TNS methods are optimally suited to assist in the design and verification of experimental protocols.

Finally, while comparably precise TNS simulations in higher dimensions can take some time to be realized, 
in the meantime we can expect that TNS help uncovering interesting features of other kinds of 
quantum field theories, even in one spatial dimension (see e.g.~\cite{lin2018lattice} in this volume),
similar to the way in which DMRG has enabled great advances in condensed matter problems.
\vspace{-.3cm}

\acknowledgments
\vspace{-.3cm}
We are thankful to the authors of \cite{Buyens2013,Buyens2015,Buyens2016} for allowing us to use their figures.
This work was partly funded by the QUANTERA project QTFLAG.
KC was partially supported by the DFG project nr. CI 236/1-1.
JIC acknowledges ERC Advanced Grant QENOCOBA under the EU Horizon 2020 program (grant
agreement 742102).
SK was supported by Perimeter Institute for Theoretical Physics. Research at Perimeter Institute is supported by the Government
of Canada through the Department of Innovation, Science and Economic Development Canada
and by the Province of Ontario through the Ministry of Research, Innovation and Science.
\vspace{-.3cm}

\bibliographystyle{JHEP}
\bibliography{TNS_and_LGT}

\end{document}